\documentclass[lettersize,journal]{IEEEtran}
\usepackage{amsmath,amsfonts}
\usepackage{algorithmic}
\usepackage{algorithm}
\usepackage{array}
\usepackage[caption=false,font=normalsize,labelfont=sf,textfont=sf]{subfig}
\usepackage{textcomp}
\usepackage{stfloats}
\usepackage{url}
\usepackage{verbatim}
\usepackage{graphicx}
\usepackage{cite}
\usepackage{xcolor}

\begin{document}

\title{An Energy-Efficient Atmospheric Plasma Jet Line Enabled by a Dielectric Microwave Anapole Source}

\author{Muhammad Rizwan Akram,~\IEEEmembership{Member,~IEEE,} and Abbas~Semnani,~\IEEEmembership{Senior Member,~IEEE}
\thanks{The authors are with the Department of Electrical Engineering and Computer Science, The University of Toledo, Toledo, Ohio 43606, USA. (email: muhammadrizwan.akram@utoledo.edu; abbas.semnani@utoledo.edu). This work was supported by the National Science Foundation (NSF) under Grant ECCS-2102100.}}

\markboth{IEEE TRANSACTIONS ON Plasma Science,~Vol.~x, No.~xx, 2025}%
{Shell \MakeLowercase{\textit{et al.}}: A Sample Article Using IEEEtran.cls for IEEE Journals}


\maketitle

\begin{abstract}
Cold atmospheric pressure plasmas are crucial for applications in medicine, agriculture, and material processing, where their interaction with ambient air generates reactive oxygen and nitrogen species for disinfection and treatment purposes. Additionally, their ability to alter carbon bonds makes them valuable for processing heat-sensitive materials and etching. Conventional resonant microwave plasma sources generate high electric fields for plasma production but are typically constrained to needle-like plasma configurations. Building on our earlier anapole plasma jet, we demonstrate a 2 cm line plasma jet for large-area surface treatment by leveraging the unique properties of the anapole source. With a uniform electric field of 10$^6$ V/m along the entire 2 cm length using just 1 W of input power, this highly efficient and compact PCB-compatible device enables stable plasma operation across a broad range of helium flow rates (1–40 slpm) and power levels (4–27 W). The anapole line plasma jet stands out for its high electron density, low operating temperature, and ease of frequency tunability, making it a promising platform for next-generation plasma applications.     
\end{abstract}

\begin{IEEEkeywords}
Anapoles, dielectric resonator, frequency tunability, plasma jet, power efficiency
\end{IEEEkeywords}

\section{Introduction}
\IEEEPARstart{P}{lasma} consists of ions and electrons in a metastable state, formed by supplying energy to a neutral gas \cite{c4conrads2000}. Various energy sources can be utilized, including electrical, thermal, and optical. Typically, the resulting plasma is weakly ionized, with an ionization rate ranging from $10^{-2}$ to 10$^{-6}$. The temperature of such weakly ionized plasma is generally low, often referred to as cold plasma. Cold plasma \cite{c1grill1994,c2rossnagel1990,c3rutscher1983} is critical for numerous applications, including medicine \cite{a27lar2015}, agriculture \cite{a31ito2018}, material processing \cite{a29pen2015}, electron accelerators \cite{c5martinez2013}, propulsion \cite{a33bet2016}, and fusion \cite{a33bet2016}. The continuous advancement of efficient plasma sources is crucial due to the broad range of industrial applications.

Regarding electrical energy, early plasma sources were developed by applying a high DC voltage to a neutral gas \cite{c6moh2002}. Later, pulsed sources \cite{c7wal2007} were introduced, followed by RF \cite{c8rai2017} and microwave energy sources \cite{c9gul2015}. A common challenge across these methods is their high power requirements, often necessitating tens of kV for DC or hundreds of watts for RF/microwave gas breakdown, which can limit their applicability. EM/microwave-based plasma sources are particularly desirable due to their $\alpha$-regime discharge, which enhances stability \cite{a40sem2016,a41sem2016}. At high frequencies, plasma behaves like an electrode, with reduced ion movement, thereby minimizing ion collisions and extending the lifespan of microwave plasma devices.

A conventional microwave device approach involves delivering microwave energy via transmission lines, such as microstrip lines and waveguides \cite{c10suzuki2015,c11suzuki2015}, or coaxial lines, to an orthogonal gas channel. However, this approach requires substantial microwave power and a complex setup, especially after plasma formation due to its quasi-conductive nature. These non-resonant approaches contrast with resonator-based microwave sources, which utilize spatial energy distribution control in confined micro-volumes to enable gas breakdown at much lower power levels, typically in the range of a few watts.

Examples of resonator-based plasma sources include split-ring resonator plasma sources \cite{a46iza2003}, coaxial transmission line resonators \cite{a48cho2010}, waveguide resonators \cite{b5sadeghfam2019, b6hong2011}, and cavity resonators \cite{a45gul2015, b7wang2012, b4heuermann2012}. Additionally, evanescent mode cavity resonators \cite{a47sem2022} have been utilized for atmospheric plasma jets. These resonators are enclosed within metallic walls, which minimize radiation losses and effectively couple energy into the gas for plasma formation. However, they require complex fabrication and assembly, making them expensive. Recently, substrate-integrated waveguide-based cavity resonators have achieved plasma formation at 15-20 W at 2.45 GHz \cite{a50zha2023}. Another prototype, a planar evanescent mode cavity resonator utilizing substrate-integrated waveguide technology \cite{a51kab2023}, operates at significantly lower input power levels, ranging from 1.5 to 3 W.

These technologies effectively generate microplasma discharges in needle-shaped plasma jets and cylindrical volumes. However, they are unsuitable for uniform large-area surface treatment, particularly in material processing applications. To the best of the author's knowledge, no RF/microwave-powered line plasma jet sources currently exist. Line plasma discharges are highly desirable for large-scale surface treatments in semiconductor industries and plasma accelerators, which have traditionally relied on laser-based plasma ignition. A high electron density is essential for these applications. While previous attempts to realize plasma lines have been made, they have primarily been enclosed plasma forms, such as those based on waveguide transmission lines \cite{c10suzuki2015,c11suzuki2015} and permanent magnets \cite{c12sakawa2004}. The enclosed plasmas are not easily accessible and have low electron densities (approximately 10$^{13}$ cm$^{-3}$), which limits their practicality.

In our previous works, we introduced a novel anapole device \cite{c13akram2024} that exploits the destructive far-field interference of electric dipoles to form a cavity without requiring metallic walls, unlike typical cavity resonators. This anapole device was further used to demonstrate a microplasma jet \cite{c14akram2023}, which requires only 1 W of input power for plasma formation while exhibiting superior properties such as high electron density, frequency tunability, and compactness. Additionally, the device is fabricated entirely from PCB, with a dielectric structure facilitating gas passage, eliminating the need for complex post-fabrication assembly.

In this paper, we successfully demonstrate a 2-cm atmospheric plasma jet line using our anapole technology. The plasma line is sustained at a low power of 4-8 W for a gas flow rate of 1-15 slpm and approximately 27 W for higher gas flow rates. Moreover, a uniform electron density exceeding 10$^{16}$ cm$^{-3}$ is observed along the plasma line. The combination of low power requirements and high electron density positions compact anapole line plasma jets as a promising technology for numerous applications across multiple disciplines.

\section{Theory and Design}

\begin{figure}[!]
\centering
\includegraphics[width=0.95\linewidth]{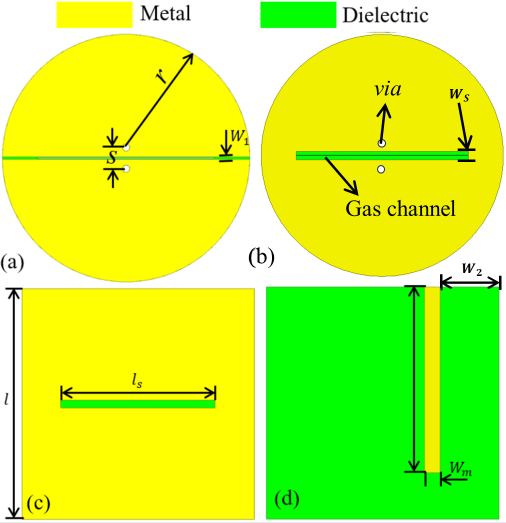}
\caption{Top (a,c) and bottom (b,d) views of the dielectric resonator and the feeding board of the anapole device with \textit{r} = 14 mm, thickness of the dielectric cylinder \textit{h} = 3.81 mm, thickness of feed board $h_2$ = 1.27 mm, $l_s$ = 20 mm, $l$ = 40 mm, $w_m$ = 2 mm, $W_2$ = 7.6 mm, $l_1$ = 24 mm, $w_s$ = 1 mm, $w_1$ = 0.3 mm, and $s$ = 1.2 mm. The cylindrical resonator is made of TMM13i with permittivity of $\epsilon_r$ = 13 and $tan\delta$ = $1.9\times10^{-3}$. The bottom board is made of TMM6 with permittivity of $\epsilon_r$ = 6 and $tan\delta$ = $2.3\times10^{-3}$.}
\label{fig1:design}
\end{figure}

The geometry of the proposed anapole plasma jet line is depicted in Fig.~\ref{fig1:design}. The device consists of two primary components: (1) a resonator and (2) a feeding board. The anapole resonator is a cylindrical disc fabricated from a Rogers TMM13 board, with a dielectric constant of $\epsilon_r = 13$ and a loss tangent of $tan(\delta) = 1.9 \times 10^{-3}$. The slot on the bottom side of the disk serves dual purposes: (1) enabling electromagnetic coupling of the resonating modes with that of the feeding board via a 50-$\Omega$ microstrip line and (2) facilitating gas flow into the disk through a valley-etched outlet measuring 2 cm by 0.1 mm.

The anapole mode in the device is achieved by incorporating two vias across the slots, connecting the top and bottom metallic patterns in a configuration that forms a split-ring resonator (SRR). The top-side metallic pattern consists of two large circles, which are disconnected from each other and shorted by their respective vias, creating a gap region within the SRR. On the bottom side, the metallic pattern is fully connected to both vias and separated by a long slot. This configuration extends the current paths over a compact device, thereby reducing the resonance frequency of the SRR.

The SRR resonance excites an additional resonance in the dielectric disk. The resulting field distribution in the disk resembles that of a cylindrical dielectric resonator antenna (CDRA) operating in the $HE_{11\delta}$ mode. The perturbation introduced by the SRR does not significantly alter the field distributions of the CDRA \cite{c15akram2018}. However, simulations performed using COMSOL indicate that these two resonances correspond to electric dipole resonances, as detailed in our previous work \cite{c13akram2024}. The radiation from these two resonances occurs in anti-phase, leading to far-field radiation cancellation without the need for metallic enclosures. Consequently, the anapole device efficiently couples all input energy while preventing radiation loss. Approximately 99.9\% of the energy is coupled into the device, with less than 5\% radiated externally. The underlying physics and detailed results on device functionality are available in our prior study \cite{c13akram2024}.

Electromagnetic energy coupling with the anapole device is ensured through a slot-coupled microstrip line. In the simulation, a TMM3 board was employed. The disk is aligned with the board so that the slot on the bottom side of the disk precisely aligns with the slot on the top side of the feeding board. On the bottom side, a 50-$\Omega$ microstrip line is used, extending perpendicularly to the slot and towards the slot edge for impedance matching. Since gas flow would disrupt the continuity of the microstrip line, a 3D-printed gas channel was designed using high-temperature resin from Formlabs. The fabrication and development details are provided in Section II.

\section{Materials and Methods}
\begin{figure}[!]
\centering
\includegraphics[width=0.85\linewidth]{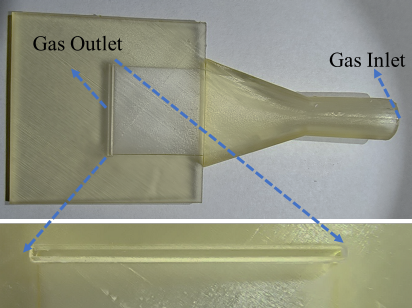}
\caption{3D-printed gas channel designed for uniform gas flow distribution, transitioning from a cylindrical inlet to a rectangular outlet. Fabricated using high-temperature resin from FormLabs.}
\label{fig2:setup}
\end{figure}

The gas channel was 3D printed using a Formlabs 3+ printer with high-temperature resin, as shown in Fig. \ref{fig2:setup}. The inlet has a cylindrical shape with an inner diameter of 5 mm, allowing direct gas feeding at the desired flow rate. The channel is tapered to ensure uniform flow towards a rectangular outlet measuring 2 cm by 0.6 mm, with a wall thickness of 0.2 mm. The outlet extends 1 mm beyond the board. Additionally, the channel serves as a feeding board for coupling electromagnetic energy.

\begin{figure}[!]
\centering
\includegraphics[width=0.9\linewidth]{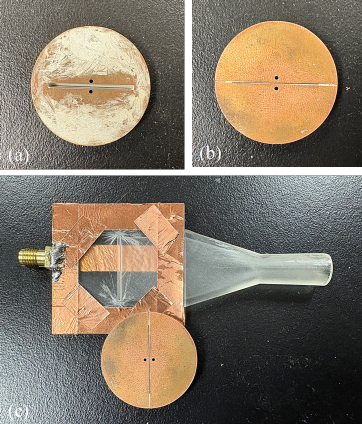}
\caption{Fabricated devices with an etched gas channel through the dielectric: (a) Top view showing a 0.1 mm × 2 cm etched gas outlet, (b) Bottom view displaying a 1 mm × 2 cm etched gas channel, and (c) Bottom board serving as both the gas flow channel and microwave feeder, integrated with the anapole resonator.}
\label{fig3:fab}
\end{figure}

\begin{figure}[!]
\centering
\includegraphics[width=0.9\linewidth]{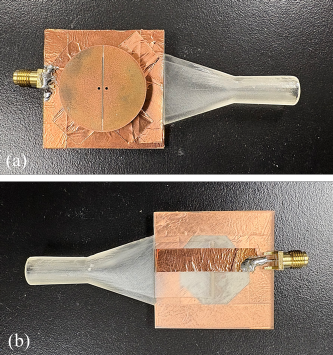}
\caption{Assembled device integrating a compact gas flow and electromagnetic feeding to the anapole resonator: (a) Top view and (b) Bottom view.}
\label{fig4:asemb}
\end{figure}

The fabricated anapole disks are shown in Fig. \ref{fig3:fab}. A slot measuring 2 cm by 1 mm was etched to a depth of 3 mm, followed by a through slot of 2 cm by 0.1 mm on the top side. The board used for electromagnetic coupling to the disk resonator measures 4 cm by 4 cm, with a thickness of 2 mm. The gas outlet is designed to fit securely on the bottom side of the anapole disk. The metallic ground on the top side of the gas channel is implemented using copper tape, while on the opposite side, a 6 mm wide microstrip line is manually positioned to ensure high impedance matching. The center pin of the 50-$\Omega$ SMA connector is soldered to the microstrip line with clearance from the ground pins on the side. Meanwhile, the ground pins on the opposite side are soldered to the ground plane, as depicted in Fig. \ref{fig4:asemb}. The final assembled anapole plasma line device is illustrated in Figs. \ref{fig4:asemb}(a) and (b).

\section{Results and Discussions}
\begin{figure*}[!]
\centering
\includegraphics[width=0.8\linewidth]{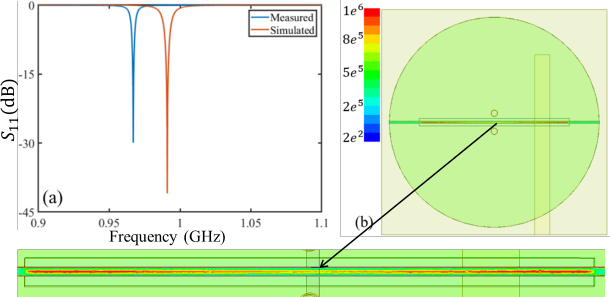}
\caption{(a) Simulated and measured reflection coefficients of the assembled device. (b) Electric field distribution at 1 W input power at the resonant frequency, demonstrating a uniform field along the 2 cm line.}
\label{fig5:s11}
\end{figure*}

To assess EM coupling to the device, its reflection characteristics were evaluated numerically using HFSS and experimentally using a calibrated Vector Network Analyzer (VNA), as shown in Fig.~\ref{fig5:s11}(a). The results indicate that nearly 100\% of the input energy is coupled to the anapole device at the experimentally measured operating frequency of 960 MHz, compared to the numerically calculated frequency of 990 MHz. The observed 30 MHz frequency shift between the fabricated and simulated results is primarily attributed to differences in the etched valleys of the disk between the simulated and experimental versions, which affect the attachment of the gas outlet. Notably, the feeding board does not influence the device's frequency response; it merely serves to transfer the input EM energy and gas flow. The electric field distribution of the anapole device is plotted in Fig.~\ref{fig5:s11}(b), showing a uniform electric field of $10^6$ V/m along the entire 2 cm line with an input power of 1 W. When helium gas is introduced into the device, the high electric field efficiently delivers microwave power, allowing for the formation of a plasma line jet.

The nature of gas flow plays a crucial role in shaping the line plasma jet and is dependent on factors such as the channel design and gas properties. This relationship is quantified using the Reynolds number, a dimensionless parameter that characterizes the flow regime \cite{a53van2015}
\begin{equation}\label{eqn1}
    Re = \rho vD_H/\eta,
\end{equation}
where $\rho$, $v$, and $\eta$ denote the fluid's density, velocity, and viscosity, respectively. In the context of gas flow within a rectangular channel, $D_H$ represents the hydraulic diameter, calculated as $D_H = 4ab/(a+b)$, where $a$ and $b$ denote the channel's width and height. For a very wide rectangular channel (i.e., $a \gg b$), $D_H$ simplifies to $2a$. A Reynolds number below 2,000 indicates laminar flow, resulting in a uniform jet; values exceeding 3,000 indicate turbulent flow, which is generally undesirable. The gas velocity is determined from the flow rate under standard conditions using
\begin{equation}\label{eqn2}
    v = 4D/\pi d^2.
\end{equation}
Here, $D$ represents the gas flow rate in slpm. Helium has a density of $\rho = 0.1634$ kg/m$^{3}$ and a viscosity of $\eta = 1.94 \times 10^{-5}$ kg/(m.s). According to (1), when $D$ is less than 40 slpm, the Reynolds number remains below 2,000, indicating a laminar flow regime.

\begin{figure}[!]
\centering
\includegraphics[width=0.95\linewidth]{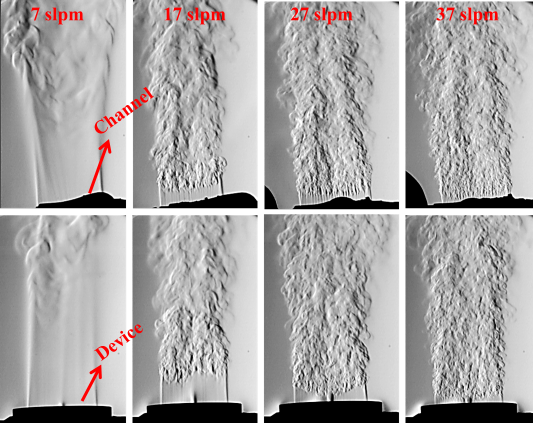}
\caption{Schlieren imaging of the gas flow for both the standalone channel and the assembled anapole line plasma device, highlighting the transition between laminar and turbulent flow regimes.}
\label{fig6:flow}
\end{figure}

To further analyze the flow dynamics, Schlieren imaging was performed at different helium flow rates, both for the gas channel alone and for the fully assembled device, as shown in Fig.~\ref{fig6:flow}. The imaging results confirm that a uniform laminar flow persists at the outlet up to a helium flow rate of 37 slpm. A slight contrast in the middle of the device is observed due to minor gas leakage through the vias.

\begin{figure*}[!]
\centering
\includegraphics[width=0.95\linewidth]{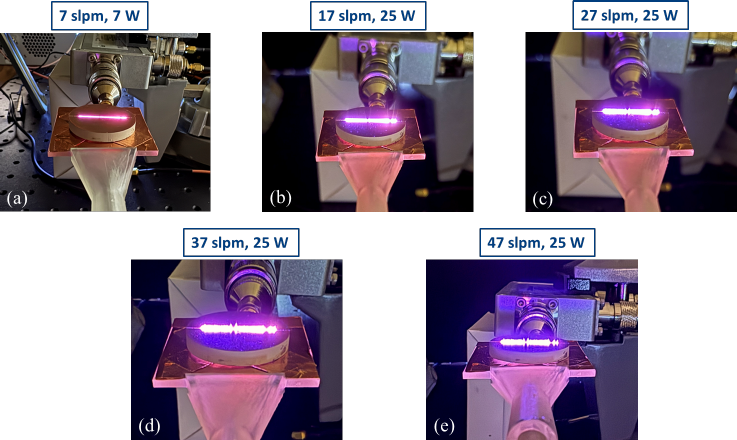}
\caption{Operation of the anapole line plasma jet at different helium flow rates, driven at an operating frequency of 960 MHz.}
\label{fig7:linePlasma}
\end{figure*}

Having characterized the flow and EM coupling mechanisms, the final anapole device was used to generate a line plasma jet. Helium gas was introduced at a controlled flow rate of 1-47 slpm using a mass flow controller. The microwave power at the resonance frequency was gradually increased, producing a uniform plasma line at 1-15 slpm with an input power of 4-8 W. However, at higher flow rates, at least 20 W was required to sustain the plasma. The resulting plasma line jets at various flow rates are shown in Fig.~\ref{fig7:linePlasma}.

At flow rates below 20 slpm, the gas flow is insufficient to eject the plasma and form a continuous line jet. As the flow rate increases to 20 slpm and above, a 4-6 mm long jet emerges. At high flow rates, the effective length of the laminar flow region decreases, limiting further elongation of the plasma jet. It should be noted that the line channel was mechanically etched using a CNC machine with a 0.1-mm diameter drill bit. Due to the narrow width, the etched channel is not entirely uniform. Laser etching techniques could be employed to achieve more precise and uniform channel structures.

\begin{figure}[!]
\centering
\includegraphics[width=0.95\linewidth]{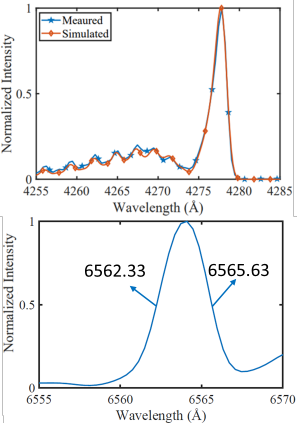}
\caption{Temperature evaluation (top) through curve fitting of experimental OES data with the simulated spectral profile from the LIFBASE database, and H-$\alpha$ spectral profile (bottom) for electron density measurement at a helium flow rate of 30 slpm and an input power of 30 W.}
\label{fig8:diagn}
\end{figure}

To characterize the electron density of the anapole line plasma jet, gas temperature measurements were performed using optical emission spectrometry. The spectral profiles of $N^{2+}$ emission were analyzed and compared with the LIFBASE database for temperature extraction. A representative temperature measurement at a flow rate of 30 slpm and an input power of 30 W is shown in Fig.~\ref{fig8:diagn}(a), yielding an extracted temperature of approximately 300 K. Measurements were conducted at three locations along the plasma line to confirm uniformity. Across the tested flow rates (1-40 slpm), the extracted temperature ranged from 300 to 350 K, ensuring operation at room temperature. The spectral profile of the H-$\alpha$ line was used to estimate the electron density ($n_e$) as
\begin{equation}
    n_e = 10^{17} \times ({\Delta \lambda_{Stark}}/1.098)^{1.47135}.
\end{equation}
Here, n$_e$ is expressed in cm$^{-3}$, and $\Delta\lambda_{Stark}$ represents the full-width half-maximum of the H-$\alpha$ spectral profile, centered at 656.3 nm.

\begin{figure}[!]
\centering
\includegraphics[width=0.95\linewidth]{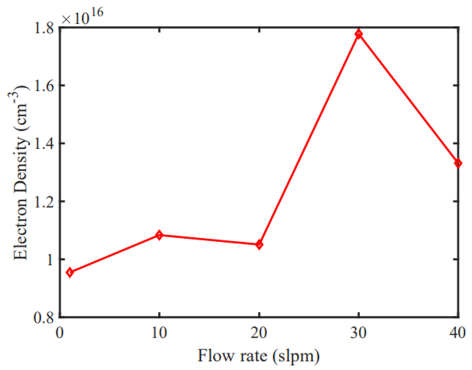}
\caption{Electron density ($n_e$) measured at various helium flow rates.}
\label{fig9:ne}
\end{figure}

The evaluated electron density was approximately $1.2\times10^{16}$ cm$^{-3}$, $1.08\times10^{16}$ cm$^{-3}$, and $9.5\times10^{15}$ cm$^{-3}$ at the three measured locations along the plasma line for a helium flow rate of 1 slpm and an input power of 5.5 W. As shown in Fig.~\ref{fig9:ne}, the electron density varies with flow rate. The highest density was observed at 30 slpm, coinciding with the optimal plasma jet formation, whereas lower densities were measured at reduced flow rates, where the plasma jet was less developed. At 40 slpm, the flow enters a turbulent regime, affecting the measurements. Overall, an average electron density of approximately $1.2 \times 10^{16}$ cm$^{-3}$ is achieved, which is at least twice the value typically obtained in conventional plasma jets employing resonant cavity approaches. 

To further characterize the cold plasma properties, the ionization rate was evaluated as
\begin{equation}
\text{Ionization rate} = \frac{n_e}{\rho_{bg}},
\end{equation}
where the background gas density, $\rho_{bg}$, is given by $\rho_{bg} = \rho_{He}/m_{He}$, with $m_{He}$ and $\rho_{He}$ representing the mass and density of helium, respectively, at a given pressure and temperature. The evaluated ionization rates are $7.4\times10^{-4}$ for a 30 slpm flow rate, $6.4\times10^{-4}$ for a 40 slpm flow rate, and approximately $5.2\times10^{-4}$ for lower flow rates, confirming the weak ionization levels desirable for cold plasma sources.

\section{Conclusion}
This work presents the first demonstration of a fully planar, compact atmospheric pressure plasma line jet operating at 960 MHz, leveraging the non-radiating physics of a metallo-dielectric anapole structure to enhance the near-electric field over a 2 cm line while suppressing far-field radiation without the need for metallic enclosures. This breakthrough paves the way for low-power, large-area plasma sources, offering a highly uniform treatment platform for industrial-scale applications in material processing, agriculture, and medicine. Notably, the achieved electron density—nearly twice that of conventional plasma jets—is uniformly distributed along the 2 cm plasma line, making it particularly advantageous for electron accelerators. The proposed technology stands out for its high electron density, compact and cost-effective design, and integrated flow mechanism, presenting significant opportunities for both fundamental research and practical applications. Given the far-reaching impact of plasma technology, these advancements are poised to drive progress across multiple scientific and industrial domains.

\section{Acknowledgment}
The authors would like to thank Dr. Omid Amili and his team for Schlieren imaging of the gas flow.

\bibliographystyle{IEEEtran}
\bibliography{ref}

\newpage

\vfill

\end{document}